\documentclass[showpacs,floatfix,preprint,amsmath,amssymb,prl]{revtex4}

\usepackage{amsmath}
\usepackage{latexsym}
\usepackage{float}
\usepackage{amssymb}
\usepackage{graphicx}

\def\ba{\begin{eqnarray}}
\def\ea{\end{eqnarray}}
\def\be{\begin{equation}}
\def\ee{\end{equation}}

\def\bm{\begin{math}}
\def\me{\end{math}}

\newcommand{\dummy}

\begin{document}
%\widetext

\title{Spinodal Decomposition in Polymer Mixtures via Surface Diffusion}

\author{J. Klein Wolterink$^1$, G.T. Barkema$^1$ and Sanjay Puri$^2$}

\affiliation{$^1$ Institute for Theoretical Physics,
Utrecht University, Leuvenlaan 4,
3584 CE Utrecht, THE NETHERLANDS. \\
$^2$ School of Physical Sciences, Jawaharlal Nehru University,
New Delhi -- 110067, INDIA.}

\begin{abstract}
We present experimental results for spinodal decomposition in
polymer mixtures of gelatin and dextran. The domain growth law is
found to be consistent with $t^{1/4}$-growth over extended
time-regimes. Similar results are obtained from lattice
simulations of a polymer mixture. This slow growth arises
due to the suppression of the bulk mobility  of polymers.
In that case, spinodal decomposition is driven by the diffusive
transport of material along domain interfaces, which gives rise to a
$t^{1/4}$-growth law.
\end{abstract}

\maketitle

A homogeneous binary mixture (AB) becomes thermodynamically unstable
when it is quenched into the miscibility gap. The subsequent evolution
of the mixture is characterized by the emergence and growth of domains
enriched in either component \cite{ab94,bf01,ao02}. This process of
{\it domain growth} or {\it coarsening} or {\it spinodal decomposition}
is of great technological and scientific importance, and is relevant in
diverse applications, e.g., the food-processing industry, metallurgy,
materials science, structure formation in the early universe, etc.
The standard experimental tools to characterize domain growth are
(a) the time-dependence of the typical domain size $\ell(t)$; and
(b) the functional form of the structure factor, which depends on
the system morphology. The domain growth law depends
critically on the mechanism which drives phase separation. For example,
diffusive transport of material through bulk domains yields the
Lifshitz-Slyozov (LS) growth law $\ell(t) \sim t^{1/3}$, whereas
diffusion along surfaces (interfaces) gives rise to a slower growth
law $\ell(t) \sim t^{1/4}$ \cite{to88,lm92,pbl97,gbp05}. For binary
fluid mixtures, the intermediate and late stages of phase separation
are driven by the advective transport of material. The relevant growth
laws (in $d=3$) are $\ell(t) \sim t$ in the {\it viscous hydrodynamic}
regime, followed by $\ell(t) \sim t^{2/3}$ in the {\it inertial
hydrodynamic} regime \cite{ab94,bf01,ao02}.

In the food-processing industry, many applications involve
domain growth in demixing polymer blends or quasi-binary
polymer solutions (i.e., two kinds of polymers in a solvent).
Polymeric phase separation has been studied experimentally in films
of polymer blends~\cite{Lauger94,Takeno98,Hayashi00,Hayashix},
in thicker samples~\cite{Wiltzius91,Kuwahara92}, and in soluted
mixtures~\cite{Steinhoff97}. The typical time-scales of decomposition
in these experiments vary from seconds \cite{Steinhoff97} to days
\cite{Kuwahara92}. To characterize the speed of the coarsening
process, one often fits a power-law to the location $q_{\rm max}$
of the maximum in the structure factor, as a function of time.
Measurements of the resulting growth exponent $\alpha$ in $q_{\rm max}
\sim t^{-\alpha}$ range from
0 to 0.45 \cite{Lauger94,Hayashi00,Hayashix}. Wilzius and Cumming
\cite{Wiltzius91} report a value of $\alpha \simeq 0.28$; Kuwahara et al
\cite{Kuwahara92} report an evolution of this exponent from $\alpha \simeq
0$ to $\alpha \simeq 0.3$; and Takeno and Hashimoto \cite{Takeno98}
measure a value between 0.25 and 0.33. Clearly, there is no consensus
on the experimental side regarding the growth law. Moreover, the dominant
mechanism for domain growth may differ in various experiments and
time-regimes, explaining the diverse exponents reported in
the literature.

In this letter, we present results from an experimental and
numerical study of spinodal decomposition in polymer mixtures in
the high-viscosity regime (i.e., without hydrodynamics). First, we will
present results from a light-scattering study of phase separation in a
mixture of gelatin and dextran, soluted in water. For deep quenches,
domain growth in this system is consistent with the growth law $\ell(t)
\sim t^{1/4}$. Next, we will present numerical results obtained from a
lattice simulation of spinodal decomposition
in a polymer mixture. Again, the
domain growth law is found to be compatible with $\ell(t) \sim t^{1/4}$
over an extended time-regime. We interpret these growth laws in the
context of phase separation driven by surface diffusion, as the bulk
mobility is drastically reduced in these polymer mixtures due to a large
free-energy barrier for releasing a component
from its concentrated phase. We conclude
this letter with a discussion of the growth exponent which characterizes
surface diffusion, and the crossover from surface to bulk diffusion.

Let us first present details of the experiments. The chemicals used for the
spinodal decomposition experiments were gelatin (fish gelatin with a
high molecular weight, Multi Products, Lot 9187, box 47) and dextran
(obtained from Sigma, industrial grade with an average molecular weight of
181 kg/mol, D-4876, Lot 41K1243). These chemicals were used without further
purification. The concentration in the final sample for gelatin was
2.46\%\  and for dextran 2.40\%\ w/w. No  salt was added to
the solution. However, azide was added
to the stock solutions to halt bacterial growth. The light-scattering
set-up used is shown in Fig.~2 of Tromp et al \cite{Tromp95}. A
He/Ne laser (633 nm) was used to generate a scattering pattern that
was projected onto a screen of tracing paper.  This is recorded
by a CCD-camera (EG \& G PARC, Model 1430P). (See also Lundell et al
\cite{Lundell05}.) The usable $q$-range was 0.3-7 $\mu$m$^{-1}$.

The gelatin-dextran mixture described above
is homogeneous at room temperature and starts to demix
at a temperature of $\simeq 8^{\circ}$C. The phase separation is clearly
visible at 6$^{\circ}$C. This is close to the gelation temperature of the
gelatin, which is estimated as 8-10$^{\circ}$C \cite{Tromp02}. We
specifically selected this mixture to obtain a high viscosity, so that
hydrodynamic effects can be ignored on the time-scales of our experiments.

\begin{figure}[htb]
\centering
\includegraphics*[width=.8\textwidth]{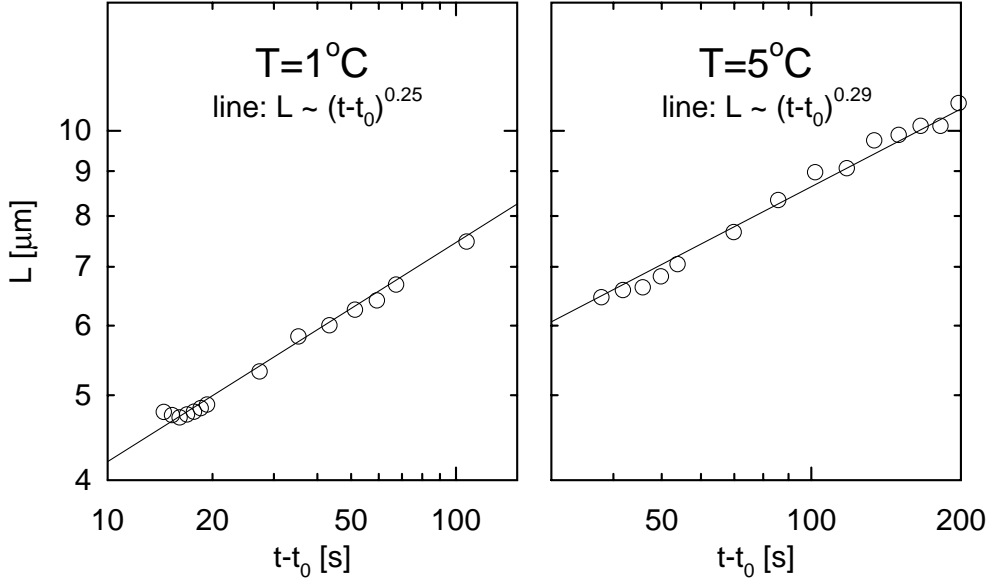}
\caption[]{\label{fig1}
Time-dependence of length scale from experiments, plotted on a log-log
scale. We study a deep quench to 1$^{\circ}$C (left frame)
and a shallow quench to 5$^{\circ}$C (right frame). The lines
denote the best linear fits to the experimental
data. The corresponding growth exponents are specified in the figure.}
\end{figure}

Starting with a homogeneous mixture,
we performed either a shallow quench to a final temperature
of 5$^{\circ}$C, or a deep quench to 1$^{\circ}$C. From then on, the
structure of the coarsening mixture is recorded at regular times with
light scattering. Specifically, we obtain the typical domain size from
$\ell(t) = 2\pi/q_{\rm max}(t)$, where $q_{\rm max}(t)$ is the location
where the spherically-averaged structure factor $S(q,t)$ reaches
its maximum. Due to practical reasons, the time $t_0$ of the onset of
phase separation is not known precisely in our experiments.
We therefore determine $t_0$
from our measurements of $S(q,t)$ as follows. Assuming that the domain
growth process is described by $q_{\rm max}(t) \sim (t-t_0)^{-\alpha}$ with
$\alpha \simeq 1/4$ or 1/3, we perform a linear least-squares fit in
which $q_{\rm max}^{-1/\alpha}$ is plotted vs. $t$; the time of zero-crossing
is then our estimate of $t_0$. Since it takes some time before the
structure factor develops a clear peak, we repeated this procedure also
for the locations $q_{\rm max}'$ and $q_{\rm max}''$ of the maxima in
$q S(q,t)$ and $q^2 S(q,t)$. This results in six estimates for $t_0$,
which we average. We use the average value of $t_0$ to make a
double-logarithmic plot of $\ell(t)$ vs. $(t-t_0)$; the result is shown in
Fig.~\ref{fig1}.

As the straight lines in the figure indicate, domain growth
after a deep quench to 1$^{\circ}$C is fitted well by $\ell(t)
\sim (t-t_0)^{0.25}$, and after a shallow quench to
5$^{\circ}$C by $\ell(t) \sim
(t-t_0)^{0.29}$. The deep quench shows convincingly a domain growth
exponent in agreement with surface diffusion. The shallow-quench
experiment is not so conclusive. The larger growth exponent
indicates that both surface and bulk diffusion contribute to
coarsening, and we are in a crossover regime between $\alpha = 1/4$
and $\alpha = 1/3$. As we will discuss shortly, the bulk mobility
is enhanced at higher temperatures where A-rich bulk domains contain
significant amounts of B, and vice versa.

We have also performed extensive simulations of spinodal decomposition
in a quasi-binary polymer mixture. The specific model that we used is a
lattice polymer model.  Polymers with $N$ bonds are described in this
model as self-avoiding and mutually-avoiding walks on a face-centered-cubic
(FCC) lattice.  The polymers move by a sequence of single-monomer jumps
to nearest-neighbor lattice sites. For monomers which are neighbors
along the same polymer, the steric interactions are lifted (and a site
can thus be occupied by two or more adjacent monomers), as this
allows enhanced reptation.  Both reptation and Rouse dynamics are
captured qualitatively in this model.
Therefore, it can be used to study dynamical
properties with some degree of realism, as long as hydrodynamic
interactions are not essential. The model is described in detail by
Heukelum and Barkema~\cite{heukelum03}.
It has previously been used to study the
fractionation of polydispersed polymer mixtures~\cite{heukelum03b},
diffusion and exchange of adsorbed polymers~\cite{joannepoluitw},
and desorption of polymers~\cite{des}.

The current study involves 128,000 polymers, each consisting of $N=50$
bonds and connecting 51 monomers. These are located on a 3-dimensional
FCC lattice with $8 \times 10^6$
sites and periodic boundary conditions. Half of the polymers
are labeled as type A, and the other half as type B.
Polymers of different type
repel each other: the total energy is equal to the number of pairs of
neighboring sites occupied by different-type polymers, multiplied by
an energy scale $J$. If an attempted Monte Carlo (MC) move increases the
total energy by $\Delta E$, the acceptance probability of this move
is $P_a=\exp(-\beta \Delta E)$, where $\beta = 1/(k_B T)$
is the inverse temperature. Moves which do not raise the total energy
are always accepted.

\begin{figure}[htb]
\centering
\includegraphics*[width=.8\textwidth]{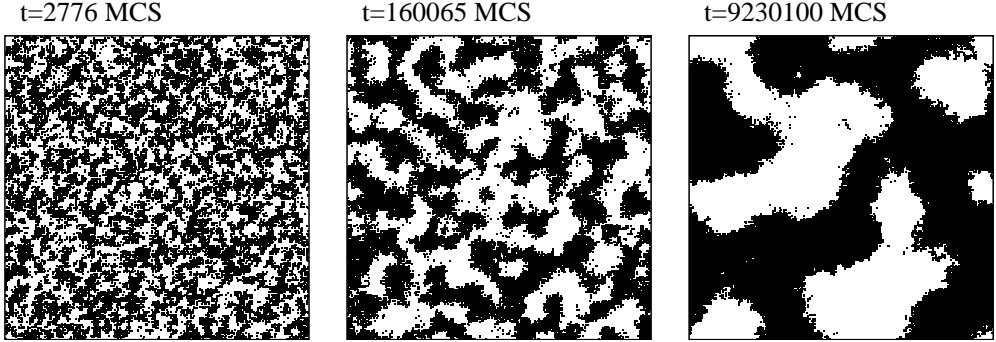}
\caption[]{\label{fig2}
Evolution pictures from simulations of spinodal decomposition in polymer
mixtures. Details of the simulation are described in the text. The quench
temperature was $\beta J=0.1$, and the system size is $200^3$. The species A is
marked in black, and the species B is not marked. The time in Monte
Carlo steps (MCS) is indicated above each snapshot.}
\end{figure}

\begin{figure}[htb]
\centering
\includegraphics*[width=.5\textwidth]{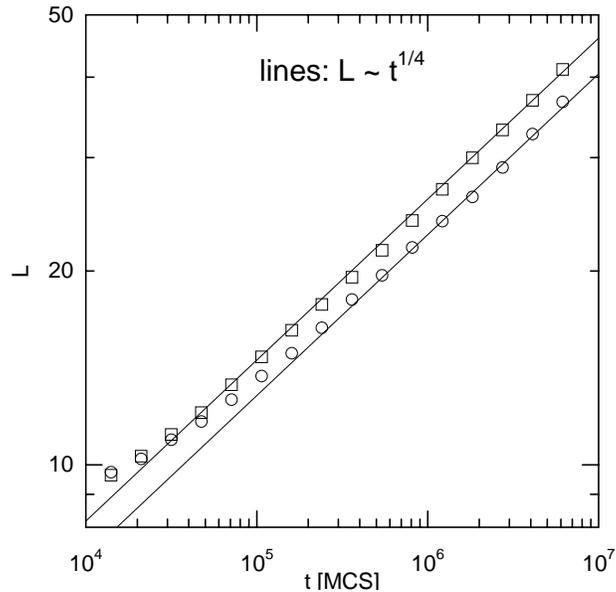}
\caption[]{\label{fig3}
Time-dependence of length scale from simulations, plotted on a log-log
scale. The time is measured in MCS. We show data for quenches to
$\beta J = 0.05$ (circles), and $\beta J = 0.1$ (squares). The
lines correspond to the growth law $\ell(t) \sim t^{1/4}$.}
\end{figure}

The simulations start from a mixed state, generated as an equilibrium
configuration at high temperature ($\beta=0$). At $t=0$, the temperature
is quenched to either $\beta J=0.05$ or $\beta J=0.1$. At these low
temperatures, the repulsive AB
interactions induce phase separation into domains
rich in either A or B. The size of these domains grows in
time -- evolution snapshots for a typical run are shown in Fig.~\ref{fig2}.
To characterize domain growth, we assign to lattice site $i$ occupied
by an A-type (B-type) polymer the value $\sigma_i=+1$ ($\sigma_i=-1$).
The domain size is defined as the
distance $\ell$ of the first zero-crossing of the spatial correlation
function of $\sigma_i$, and we measure it at various times. The
time-dependence of $\ell (t)$ is shown in Fig.~\ref{fig3}. After an
initial transient regime, both data sets (for $\beta J = 0.05,0.1$)
are consistent with the growth law $\ell(t) \sim t^{1/4}$ over extended
time-regimes, viz., $\sim 2$ decades for $\beta J = 0.05$ (denoted by
circles), and $\sim 2.5$ decades for $\beta J = 0.1$ (denoted by squares).

The above experimental and numerical results can be understood in
the following theoretical framework. The kinetics of phase separation
is described by the Cahn-Hilliard (CH) model \cite{ab94,bf01,ao02}:
\begin{equation}
\label{ch}
\frac{\partial}{\partial t} \psi (\vec{r},t) =
\vec{\nabla} \cdot \left[ D(\psi) \vec{\nabla} \left(
\frac{\delta F}{\delta \psi} \right) \right],
\end{equation}
where $\psi (\vec{r},t)$ is the order parameter at space point $\vec{r}$
and time $t$. Typically, $\psi (\vec{r},t) \simeq \rho_A (\vec{r},t)
- \rho_B (\vec{r},t)$, where $\rho_A$ and $\rho_B$ denote the local
densities of species A and B. We have neglected thermal fluctuations in
Eq.~(\ref{ch}), and considered the general case of a $\psi$-dependent
mobility $D(\psi)$ \cite{lbm75,ki78,pbd92}.  Further, $F[\psi]$ denotes
the Helmholtz potential, which is usually taken to have the $\psi^4$-form:
$F[\psi] \simeq \int d\vec{r} \left[ - \psi^2/2 + \psi^4/4 + (\vec{\nabla}
\psi)^2/2 \right]$, where we have used the order-parameter scale and
the bulk correlation length to express $F[\psi]$ in dimensionless units.

Let us consider a situation where the mobility at the interfaces
$(\psi=0)$ is $D_s$, and that in the bulk $(\psi = \pm 1)$ is $D_b$, with
$D_b \leq D_s$. This difference in interfacial and bulk mobilities can
result from low-temperature dynamics, where there is a high energy barrier
for an A-atom to enter a B-rich bulk domain and vice versa.
Alternatively, physical
processes like glass formation \cite{sj97} or gelation \cite{sbs93,sp01}
can also result in a drastic reduction in bulk mobility. A simple
functional form for the mobility in the above case is $D(\psi) = D_s
(1-\alpha \psi^2),~\alpha = 1 - D_b/D_s$.  Notice that $0 \leq \alpha
\leq 1$ for $D_b \leq D_s$.  Thus, Eq.~(\ref{ch}) can be written as
\ba
\label{chd}
\frac{\partial}{\partial t} \psi (\vec{r},t) = \vec{\nabla} \cdot
\left[ \left( 1 - \alpha \psi^2 \right) \vec{\nabla}
\left( -\psi + \psi^3 - \nabla^2 \psi \right) \right] ,
\ea
where we have absorbed $D_s$ into the time-scale.

Equation~(\ref{chd}) can be decomposed as \cite{lm92}
\ba
\label{chsplit}
\frac{\partial}{\partial t} \psi (\vec{r},t) =
(1-\alpha) \nabla^2 (-\psi + \psi^3 - \nabla^2 \psi) +
\alpha \vec{\nabla} \cdot \left[ \left( 1 - \psi^2 \right) \vec{\nabla}
\left( -\psi + \psi^3 - \nabla^2 \psi \right) \right] ,
\ea
where the first term on the RHS corresponds to bulk diffusion. This term
is absent in the limit $\alpha = 1$ ($D_b=0$). The second term on the
RHS corresponds to surface diffusion. It is only relevant at interfaces,
where $\psi \simeq 0$. The location of the interfaces $\vec{r}_i(t)$
is defined by the zeros of the order-parameter field, $\psi \left[
\vec{r}_i(t),t \right] = 0$.  Let us focus on a particular interface in
the $d$-dimensional case.  We denote the normal coordinate as $n$ and the
interfacial coordinates as $\vec{a}$ [with dimensionality $(d-1)$]. Then,
the normal velocity $v_n (\vec{a},t)$ obeys the integro-differential
equation \cite{to88,lm92}:
\ba
\label{int}
4 \int d \vec{a'} G[\vec{r}_i(\vec{a}), \vec{r}_i(\vec{a'})]
v_n (\vec{a'},t) \simeq (1-\alpha) \sigma K (\vec{a},t) + 
4 \alpha \int d \vec{a'}
G[\vec{r}_i (\vec{a}), \vec{r}_i (\vec{a'})] \nabla^2 K (\vec{a'},t) .
\ea
Here, $K (\vec{a},t)$ is the local curvature at point $\vec{a}$ on the
interface, and $\sigma$ is the surface tension. The Green's function
$G(\vec{x},\vec{x'})$ is obtained from $-\nabla^2 G(\vec{x},\vec{x'})
= \delta (\vec{x} - \vec{x'})$.

A dimensional analysis of Eq.~(\ref{int}) yields the
growth laws due to surface and bulk diffusion. We identify the scales of
various quantities in Eq.~(\ref{int}) as
\ba
&& [d\vec{a}] \sim \ell^{d-1}, \quad [G] \sim \ell^{2-d}, \nonumber \\
&& [v_n] \sim \frac{d\ell}{dt}, \quad [K] \sim \ell^{-1}.
\ea
This yields the crossover behavior of the length scale as
\ba
\ell(t) & \sim & (\alpha t)^{1/4} , \quad t \ll t_c , \nonumber \\
& \sim & [(1-\alpha) \sigma t]^{1/3}, \quad t \gg t_c ,
\ea
where $t_c \sim \alpha^3/[(1-\alpha)^4 \sigma^4]$. Notice that the
asymptotic regime obeys the LS growth law, which is characteristic
of growth driven by bulk diffusion. However, the crossover to
the LS regime can be strongly delayed if the bulk mobility is
suppressed relative to the surface mobility, i.e., $D_b/D_s \rightarrow
0$ or $\alpha \rightarrow 1$.

In conclusion, we have presented results from both experiments and
simulations of spinodal decomposition in polymer mixtures. In both cases,
we find an extended time-regime where the domain growth law is compatible
with $\ell(t) \sim t^{1/4}$. This slower growth law arises because there
is a strong suppression of mobility in bulk domains due to a large energy
barrier to dissolve in the polymer-poor phase. Therefore, domain growth
is primarily driven by diffusive transport along interfacial regions,
which gives rise to a $t^{1/4}$-growth law. Understanding the mechanism of
demixing is not only important for identifying the domain growth exponent,
it is also a pre-requisite for obtaining control over the structure
formation. We hope that the results presented here will stimulate further
experimental and numerical interest in this fascinating problem. \\

\noindent Acknowledgment: We would like to thank Hans Tromp from NIZO
food research for his kind hospitality, and Arjan Visser for
preliminary simulations.

\end{document}